\documentclass[conference]{IEEEtran}

	\usepackage{chngpage}	
	\usepackage{float}	
	\usepackage[hyphens]{url}
	\usepackage[bookmarks=false]{hyperref}
	\usepackage{ctable}
	\usepackage[flushleft]{threeparttable}
	\usepackage{comment}
	\usepackage{listings}
	\usepackage{xcolor} 
	\usepackage{algorithm2e}
	\usepackage{listings}
  \usepackage{graphicx}

	\usepackage[font=bf,skip=\baselineskip]{caption}

\lstdefinestyle{turtle}{%
    morekeywords={a, @prefix},
    morecomment=[s][\textrm]{<}{>},
    morecomment=[s][\textit]{"}{"},
}

\newcommand{\furl}[1]{\footnote{\scriptsize \url{#1}}} 

\begin{document}
%
\title{Towards a Semantic Administrative Shell for Industry 4.0 Components}

\author{
\IEEEauthorblockN{Irl\'an Grangel-Gonz\'alez\IEEEauthorrefmark{1}, Lavdim Halilaj\IEEEauthorrefmark{1}, G{\"o}khan Coskun\IEEEauthorrefmark{1}, S{\"o}ren Auer\IEEEauthorrefmark{1}, Diego Collarana\IEEEauthorrefmark{1}, Michael Hoffmeister\IEEEauthorrefmark{2}}
\IEEEauthorblockA{\IEEEauthorrefmark{1}Enterprise Information Systems, University of Bonn, Germany
\\\{grangel,halilaj,coskun,auer,collaran\}@cs.uni-bonn.de}
\IEEEauthorblockA{\IEEEauthorrefmark{2}Festo AG \& Co. KG, Germany\\\{michael.hoffmeister@festo.com\}}
}

\maketitle

\begin{abstract}
In the engineering and manufacturing domain, there is currently an atmosphere of departure to a new era of digitized production.
In different regions, initiatives in these directions are known under different names, such as \emph{industrie du futur} in France, \emph{industrial internet} in the US or \emph{Industrie 4.0} in Germany.
While the vision of digitizing production and manufacturing gained much traction lately, it is still relatively unclear how this vision can actually be implemented with concrete standards and technologies.
Within the German Industry 4.0 initiative, the concept of an Administrative Shell was devised to respond to these requirements.
The Administrative Shell is planned to provide a digital representation of all information being available about and from an object which can be a hardware system or a software platform.
In this paper, we present an approach to develop such a digital representation based on semantic knowledge representation formalisms such as RDF, RDF Schema and OWL.
We present our concept of a \emph{Semantic I4.0 Component} which addresses the communication and comprehension challenges in Industry 4.0 scenarios using semantic technologies. 
Our approach is illustrated with a concrete example showing its benefits in a real-world use case.
\end{abstract}

\IEEEpeerreviewmaketitle

\section{Introduction}

The dynamic of today's world imposes new challenges to the enterprises. 
The globalization, the ubiquitous presence of the internet and the development of hardware systems are some of the technological improvements that provoke changes everywhere.   
In the engineering and manufacturing domain, there is currently an atmosphere of departure to a new era of digitized production.
In different regions, initiatives in these directions are known under different names, such as \emph{industrie du futur} in France, \emph{industrial internet} in the US or \emph{Industrie 4.0} in Germany.

Industry 4.0 (I4.0) is a term coined in Germany to refer to the fourth industrial revolution.
This is understood as the application of concepts such as Internet of Things (IoS), Cyber-physical Systems (CPS), the Internet of Services (IoS) and data-driven architectures in the real industry.
With approximately the similar meaning, in North America, the term Industrial Internet has been created. 
This term is very similar to I4.0, but the application is broader than industrial production.
Other areas are included, for instance, smart electrical grids~\cite{drath2014industrie}.

With the goal to develop the Industry 4.0 vision, CPS are of paramount importance. 
CPS integrate physical and software processes~\cite{lee2008cyber}.
In order to do so, they use various types of available data, digital communication facilities, and services~\cite{mikusz2014towards}.

While the vision of digitizing production and manufacturing gained much traction lately, it is still relatively unclear how this vision can be actually be implemented with concrete standards and technologies.
The physical network connection problem is meanwhile largely solved using technologies such as \emph{Profibus/Profinet}~\cite{808018} and \emph{OPC-UA}~\cite{mahnke2009opc}.
However, the much more challenging problem is to make smart industrial devices able to communicate and \emph{understand} each other as a prerequisite for cooperation scenarios.
To address this problem, we need techniques and standards for representing and exchanging information, data and knowledge between devices participating in manufacturing and production processes.
Such standards must be flexible to accommodate new features, usage scenarios, cover multiple domains, device categories, and bridge organizational boundaries.
Most importantly, they must be able to evolve seamlessly over time to facilitate the swift realization of new features and scenarios as they become apparent.

Within the Industry 4.0 initiative, the concept of an \textit{Administrative Shell} was devised to respond to these requirements.
The Administrative Shell is planned to provide a digital representation of all information (and services) being available about and from a physical manufacturing component.
In this article, we present an approach to develop such a digital representation based on semantic knowledge representation formalisms such as RDF, RDF-Schema and OWL.
The advantages of such an RDF-based approach are:
\begin{itemize}
	\item \emph{Identification.} The use of URI/IRIs provides an decentralized, holistic, extensible global identification scheme for all relevant entities either physical or abstract
	\item \emph{Integration} The simple, but at the same time expressive statement-centric RDF data model enables the representation of facts, raw data, schema, provenance and meta-data information in a unified manner.
	\item \emph{Coherence} Existing and new taxonomies, vocabularies, and ontologies can be mixed and meshed to represent information about various domains, application scenarios, organizations etc.
\end{itemize}

The remainder of this paper is structured as follows. 
An overview of background information and descriptions of the relevant terminology for our approach is provided in \autoref{background}.
A comprehensive list of challenges aggregated from the current state of the art and our ongoing work on is presented in \autoref{challenges}. 
In \autoref{semI40component}, we present \emph{Semantic I4.0 Component} which is our approach to addressing the challenges using semantic web technologies. 
In addition, a concrete example is given in \autoref{usecase} which shows the benefits of our approach in real world use case.
We provide an overview about related work in \autoref{relatedWork}.
The conclusion and an outlook to future work are presented in \autoref{conclFutureWork}.

\section{Background}
\label{background}
This section describes several concepts and terminology that are relevant for our approach.

\subsection{RAMI Model}
\begin{figure}[b]
	\centering
		\includegraphics[width=1.00\columnwidth]{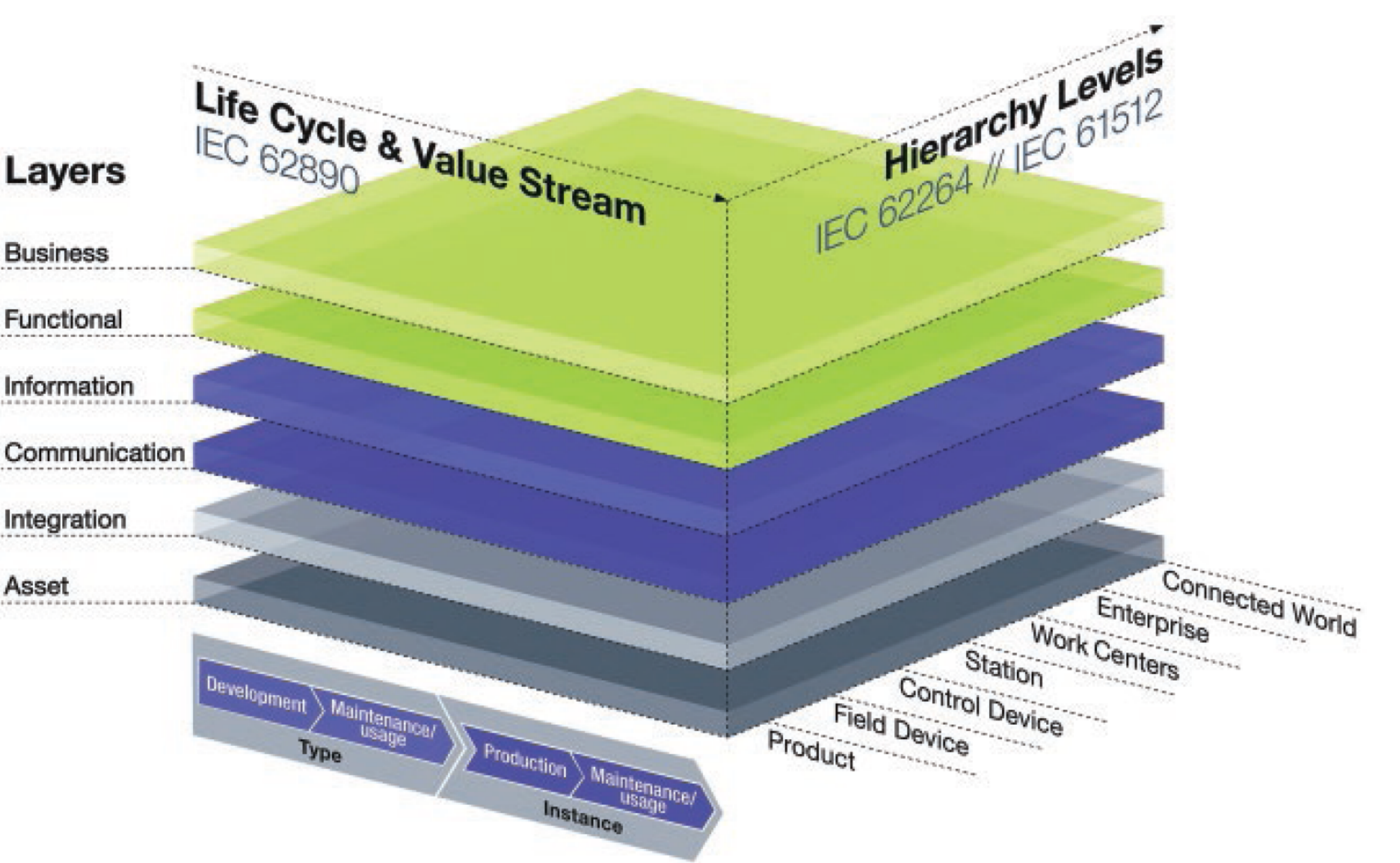}
	\caption{Reference architecture model for Industrie 4.0 (RAMI 4.0) comprising the three dimensions layers, life-cycle and hierarchy levels (from \cite{ramiModel}).}
	\label{fig:rami}
\end{figure}

Several German institutions and associations worked in close cooperation to define a reference model for Industry 4.0.
The result is the \emph{Reference Architecture Model for Industry 4.0} (RAMI 4.0), that describes fundamental aspects of the Industry 4.0~\cite{ramiModel}.
It illustrates the connection between IT, manufacturers/plants and product life cycle through a three-dimensional space.
Each dimension shows a particular part of these worlds divided into different layers as depicted in \autoref{fig:rami}.
Left vertical axis represents IT perspective which is comprised of various layers such as business, functional, information, etc. 
These layers corresponds to the IT way of thinking where complex projects are decomposed into smaller manageable parts.
In the left hand horizontal axis is displayed the product life cycle where \emph{Type} and \emph{Instance} are distinguished as two main concepts.
The model allows the representation of the data gathered during the entire life cycle.
Along with the right hand horizontal axis the location of the functionalities and responsibilities are given in the hierarchical organization.
The model broadens the hierarchical levels of IEC 62264 \furl{http://www.iso.org/iso/catalogue_detail.htm?csnumber=57308} by adding the \emph{Product} or a \emph{workpiece level} at the bottom, and the \emph{Connected World} goes beyond the boundaries of the individual factory at the top.
In addition, the reference architecture model allows the description and implementation of highly flexible concepts.
This leverages the transition process of current manufacturing systems to Industry 4.0 by providing an easy step by step migration environment.

\subsection{Industry 4.0 Component}
\begin{figure}[tb]
  \centering    
  \includegraphics[width=0.6\linewidth]{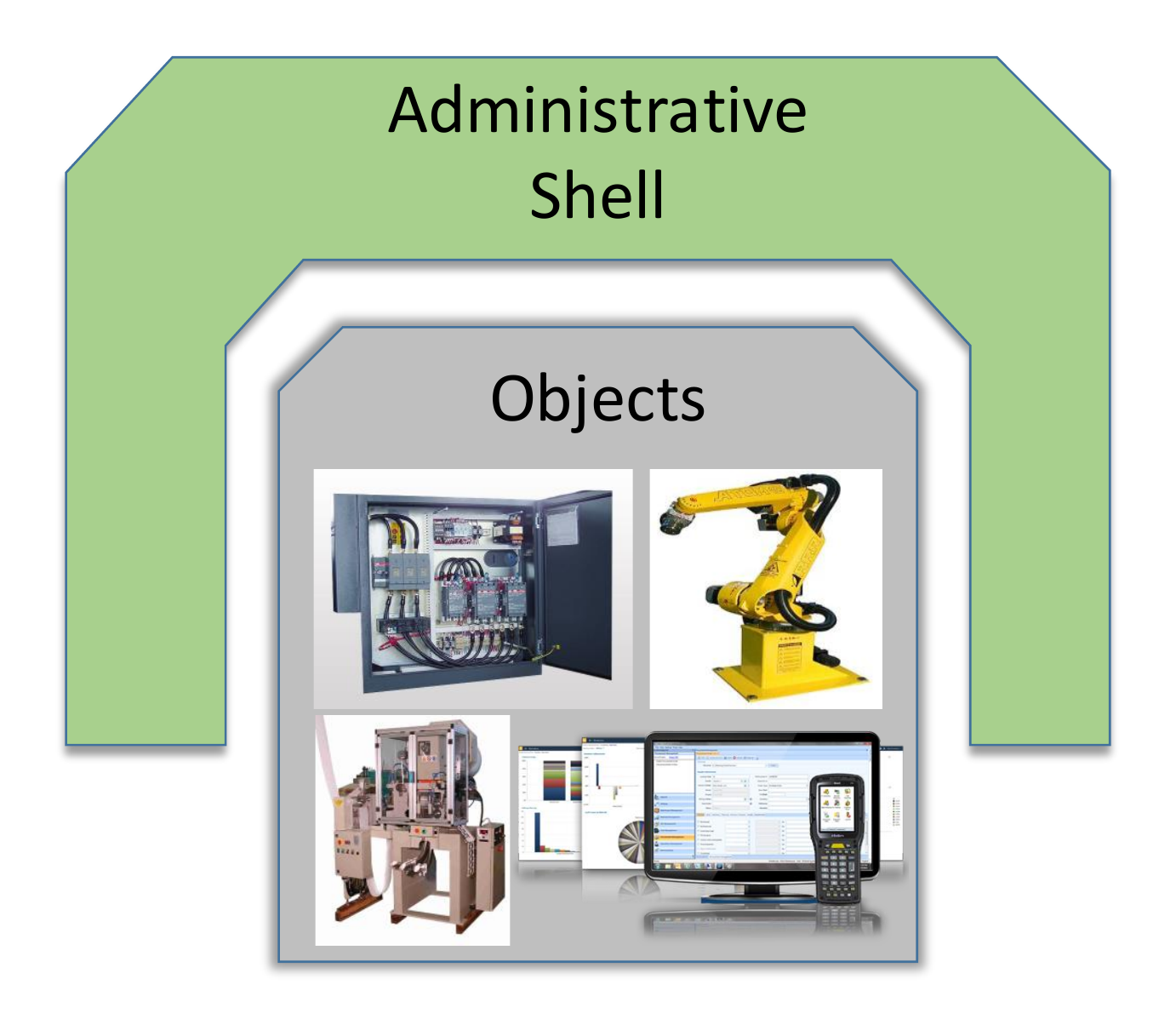}
  \caption{Concept of an Industry 4.0 Component consisting of the object and its Administrative Shell.}
  \label{fig:I40Component}
\end{figure}

A component is a basic concept in Industry 4.0.
As defined in~\cite{ramiModel} an I4.0 component constitutes a specific case of a CPS.
It is used as a model for representing the properties of CPS, for instance, real objects in a production environment connected with virtual objects and processes. 
An I4.0 component can be a production system, an individual machine, or an assembly inside a machine. 
It is comprised of two foundational elements: object and Administrative Shell.
Every object or entity that is surrounded by an Administrative Shell is described as I4.0 component.
\autoref{fig:I40Component} shows an example of such a component.
Additionally, these objects have at least the capability of passive communication.
As a result, a flexible framework for data description and provisioning is established.
In the following, these elements are presented in detail.

\subsection{Object}
In \cite{ramiModel}, the term \emph{object} is used to refer each individual physically or non-physically part.
An object can be an entire machine, an automation component or a software platform.
From a time perspective, it could be a legacy system or a new system developed by modern techniques and technologies.
The industry should be able to integrate and benefit from these objects, independent of the type and time that they belong.

\subsection{Administrative Shell}
\label{AdminShell}

The Administrative Shell is used to store all important data of an object which can be hardware or software. 
It creates benefits for all participants in networked manufacturing. 
Consequently, a consistent way of managing data along with various functions and services for data manipulation and publication is provided. 
Some of the benefits are presented in detail as below~\cite{ramiModel}:

\paragraph{Management of Data} 
The Administrative Shell provides mechanisms to manage a large amount of data and information generated by manufacturers or participants.
For instance, it stores and manages the information related to configuration, maintenance or connectivity with other devices.

\paragraph{Functions} 
Different functions such as configuration, operation, maintenance, complex algorithms for business logic are provided by the Administrative Shell.
These functions facilitate the interactivity between the I4.0 component and other actors including human users.

\paragraph{Services} 
Although the information of a component is stored only once, they can be used beyond the boundaries of the component, within enterprise networks and in the cloud as well.
The advantage is that the information can be provided via IT services to any user or in any use case.

\paragraph{Integration} 
The Administrative Shell, in combination with communication protocols, offers the possibility of easy integration of each I4.0 component.

\paragraph{Modularity} 
Each specific part of an object should be able to store information in Administrative Shell.
This ensures that all information are saved and is ready to be used in further analysis.
Consequently, the Industry 4.0 is applied successfully in a networked manufacturing.

\section{Challenges}
\label{challenges}

Developing the vision of Industry 4.0 raises new challenges.
These challenges become even harder since participants in a networked manufacturing apply different policies.
This section describes some of the challenges which are more critical for I4.0 in general and I4.0 component in particular.
In the following, these requirements are presented in detail.

\subsubsection{Interoperability (Ch1)}
On the one hand, I4.0 vision includes new ways of managing data, machines and components. 
On the other hand, the enterprises need to maintain a huge amount of legacy systems with their corresponding existing data. 
Commonly, this data is in different formats (e.g. plain text, DBMS, XML, etc.).
The new data and new formats have to coexist with the old ones. 

\subsubsection{Global unique identification (Ch2)}
Enabling intercommunication among I4.0 components and the environment over the Internet is a big challenge. 
In addition to this, there should be a linking mechanism between the I4.0 components and the generated information~\cite{framling2007requirements}. 
Therefore, addressing this challenge is of paramount importance in order to realize the vision of I4.0.

\subsubsection{Data availability (Ch3)}
Another challenge is the availability of the data beyond the boundaries of the manufacturers and across different hierarchy levels.
This challenge becomes even harder when various policy rules from manufacturers are applied.
I4.0 components will communicate with each other and interact with the environment through exchanging the data generated from different sensors and react to the events by triggering actions with the aim of controlling the physical world~\cite{vermesan2011internet}. 
Therefore, sharing the generated data between participants~\cite{abu2013data} is a key factor in the Industry 4.0.

\subsubsection{Standardization compliance (Ch4)}
The Standardization process is an important step toward the realization of I4.0.
Several standards to deal with different layers in the enterprises exist nowadays.
For instance, \emph{AutomationML}~\cite{drath2010datenaustausch}, \emph{Profibus}~\cite{IEC61851} and \emph{OPC-UA}~\cite{mahnke2009opc,journals/at/EnsteM11} are just some of the examples of the mentioned standards. 
The core idea of all this effort is to provide a detailed description of the components in the manufacturing process.
The production process constantly generates different components and the standards need to reflect this dynamically.   
As a result, the standards grow in size and number, making this interoperability between them a problem to solve. 

\subsubsection{Integration (Ch5)}
Highly dynamic environment is one of the key obstacles to the establishment of the vision of I4.0.
The complexity of horizontal and vertical integration of the I4.0 components is drastically increased with the fluctuating number of the participants.
Self-sense, self-configuration and self-integration are some concepts used to describe autonomous interaction of a component with the environment in a networked manufacturing.
Following the principles from Reconfigurable Manufacturing Systems (RMS), adding, removing, replacing or rearranging the components must not affect the production process~\cite{abele2007mechanical}. 
Thus, developing a consistent data model is a crucial factor that facilitates the integration of I4.0 components in the changing environments.

\subsubsection{Multilinguality (Ch6)}
In order to achieve a wide range of applicability to different cultures and communities~\cite{hillier2003role}, the I4.0 should be able to support localization (and internationalization) of the generated information.
This will decrease the learning curve and allow easier and faster adoption of the Industry 4.0 in real production environment.

\section{Semantic I4.0 Component}
\label{semI40component}

\subsection{Addressing the challenges with the semantic approach}
\begin{figure*}[tb]
  \centering    
  \includegraphics[width=.8\linewidth]{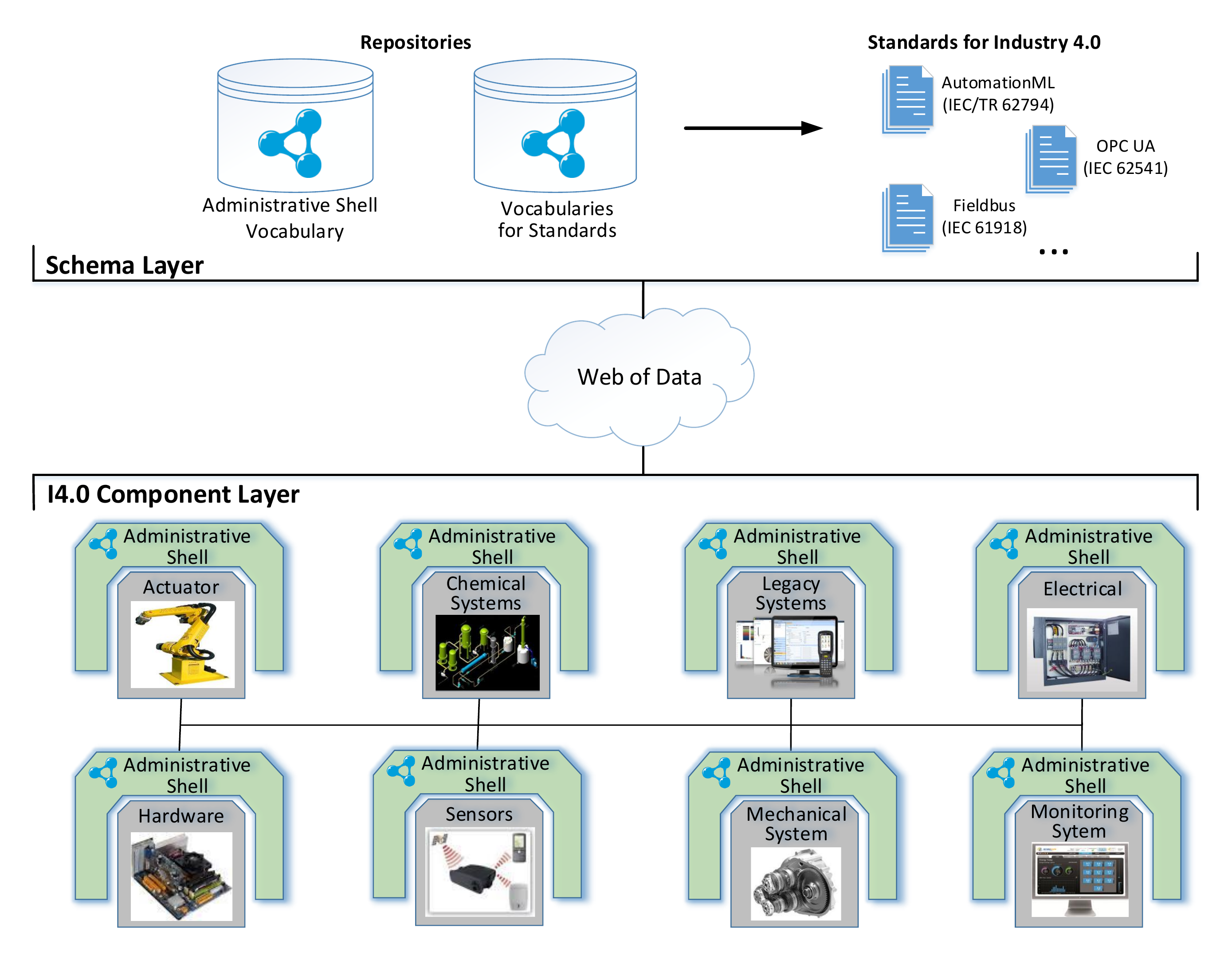}
  \caption{The Semantic I4.0 Component architecture comprising vocabularies and RDF representations of relevant standards for representing information about a wide range of relevant components.}
  \label{fig:i40Architecture}
\end{figure*}

It is widely accepted that semantics technologies play a crucial role regarding the management of things, devices and services~\cite{thoma2014managing,wahlster2014semantic}. 
Moreover, \cite{ramiModel} recognizes as a requirement that I4.0 components and their contents should follow a common semantic model.
Therefore, we propose a semantic approach to address the challenges presented in \autoref{challenges}. 
This approach is developed using fine-grained principles from Semantic Web and Linked Data.
\autoref{fig:i40Architecture} depicts our proposal to add a semantic layer to the Administrative Shell. 

\subsubsection{Interoperability}
To meet the interoperability demand, RDF and Linked Data have proven to be a successful way to integrate different types of data~\cite{langegger2008semantic,bizer2012ldif,graube2011linked}.
Embedded in the semantic for Administrative Shell we propose RDF as a middle layer to support the interoperability between the data of the legacy systems and the data generated by the I4.0 component.
We aim to establish RDF as a \emph{lingua franca} for data interoperability in the I4.0 landscape. 

\subsubsection{Global unique identification}
Identification of each I4.0 component by using global unique identifier ensures entity disambiguation and retrievable~\cite{bandyopadhyay2011internet}.
According to linked data principles~\cite{berners2009linked}, HTTP URIs should be used for naming things.
Following the above-mentioned principle, we propose that each I4.0 component should be identified by an HTTP URI.
By doing so, a decentralized, holistic and extensible global unique identification scheme for I4.0 components is established.
As a consequence, we will have derefenceable I4.0 components which are able to self-locate and communicate with each other.
\autoref{lst:GlobalID} presents our proposal for identifying the I4.0 components.
In addition, it shows that identification capabilities can be extended by various existing vocabularies that provides adequate means.
This example uses the term \emph{identifier} from \emph{Dublin Core Vocabulary}\furl{http://dublincore.org/documents/dcmi-terms/\#H1} to achieve a  reference to the resource which is unambiguous within a given context.

\begin{lstlisting}[style=turtle, caption={Global ID with RDF}, label={lst:GlobalID}]
 @prefix i40c: <http://purl.org/eis/i40c/>.
 @prefix rdf:  <http://www.w3.org/1999/02/22-rdf-syntax-ns#>.
 @prefix rdfs: <http://www.w3.org/2000/01/rdf-schema#> .
 @prefix dcterms: <http://purl.org/dc/terms/>.
 
 #Class Definition
 i40c:Actuator rdfs:subClassOf     i40c:Component ;
               rdfs:comment        "Actuator is ...";
      		   rdfs:label          "Actuator".

 #Instance Definition
 i40c:ActuatorAAA001 	a          i40c:Actuator; 
     		   dcterms:identifier  "AAA001" ;
               rdfs:label          "Actuator ID AAA001".
\end{lstlisting}

\subsubsection{Data availability}
The benefits of employing RDF as the standard for representation of the data are twofold.
Firstly, various data serialization formats are easy to be generated and transmitted over the network.
Secondly, using SPARQL\furl{http://www.w3.org/TR/rdf-sparql-query/}, as a W3C Recommendation for an RDF query language, it is possible to make data available through a standard interface. 
RDF representation of the data can be created on the fly, even if they are stored in relational databases or other data formats~\cite{conf/esws/QuilitzL08}.
By doing so, our approach enables data sharing between legacy systems and other participants in a networked manufacturing as well.

\subsubsection{Standardization compliance}
\begin{figure}[tb]  
\centering
  \includegraphics[width=.9\linewidth]{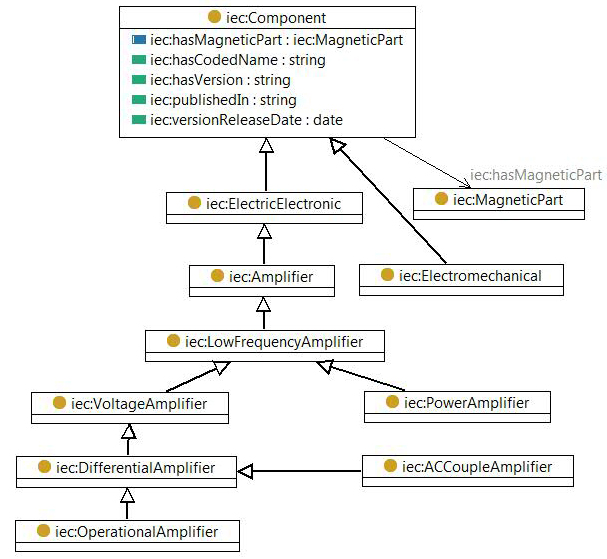}
  \caption{Overview of the class Component Hierarchy in the IEC CDD Vocabulary.}
  \label{fig:IECVoc}
\end{figure}

Following the idea of employing RDF as a lingua franca for data integration, we propose to translate existing standards into RDF vocabularies and SKOS thesauri. 
The interoperability between standards can thus be managed through the integration of the respective vocabularies. 
In addition, these vocabularies are also connected with the Administrative Shell data (cf \autoref{fig:i40Architecture}). 
As an example, we created an RDF vocabulary for the \emph{IEC 61360 - Common Data Dictionary} (IEC CDD)\furl{http://std.iec.ch/cdd/iec61360/iec61360.nsf/}.
IEC CDD is a common repository of concepts for all electrotechnical domains based on the methodology and the information model of IEC 61360.
It provides a widely accepted terminology and definitions based on accepted sources such as IEC standards as well as international and industry standards.
It contains four major concepts: \texttt{Component}, \texttt{Material}, \texttt{Feature} and \texttt{Geometry}.
\texttt{Component} describes an industrial product which serves a specific function and which in a given context is considered not to be decomposable or physically divisible and is intended for use in a higher-order assembled product.
\texttt{Component} is represented by an \texttt{Object} in RAMI Model, and when it is surrounded by the Administrative Shell forms and I4.0 component.
\autoref{fig:IECVoc} depicts part of the hierarchy of the \texttt{iec:Component} class. 

\subsubsection{Integration}
Running on completely unified and consistent data model facilitates the integration of I4.0 components.
Newly added components need a shorter time for the integration process.
Other peers will be aware of new peer and the way of communication with it by simply synchronizing with the latest version of vocabulary.
The Vocabulary contains all necessary information for interaction and data exchanging between peers in a networked manufacturing.

\subsubsection{Multilinguality}
Since various communities across the world will interact with I4.0 components, it is very important that they will receive terms in their own language.
The semantic web technologies enable implementation of multilinguality in a very straightforward manner.
This will remain valid even for the newly introduced languages or concepts.
\autoref{lst:Multilinguality} depicts an example of our approach for multilinguality.
In this example is modeled an actuator with the adequate translation for \emph{rdfs:label} and \emph{rdfs:comment} into respective languages, \emph{English} and \emph{German}.

\begin{lstlisting}[style=turtle, caption={Multilinguality}, label={lst:Multilinguality}]
 @prefix i40c: <http://purl.org/eis/i40c/>.
 @prefix rdf:  <http://www.w3.org/1999/02/22-rdf-syntax-ns#>.
 @prefix rdfs: <http://www.w3.org/2000/01/rdf-schema#> .

 i40c:ActuatorAAA001   a  i40c:Actuator; 
      dcterms:identifier  "AAA001" ;
      rdfs:label          "Actuator with ID AAA001"@en;
      rdfs:comment        "Actuator is ..."@en;
      rdfs:label          "Aktor mit ID AAA001"@de;
      rdfs:comment        "Aktor ist ..."@de.    
\end{lstlisting}

\subsection{The Administrative Shell Vocabulary}

\begin{figure*}[tb]
  \centering    
  \includegraphics[width=0.97\linewidth]{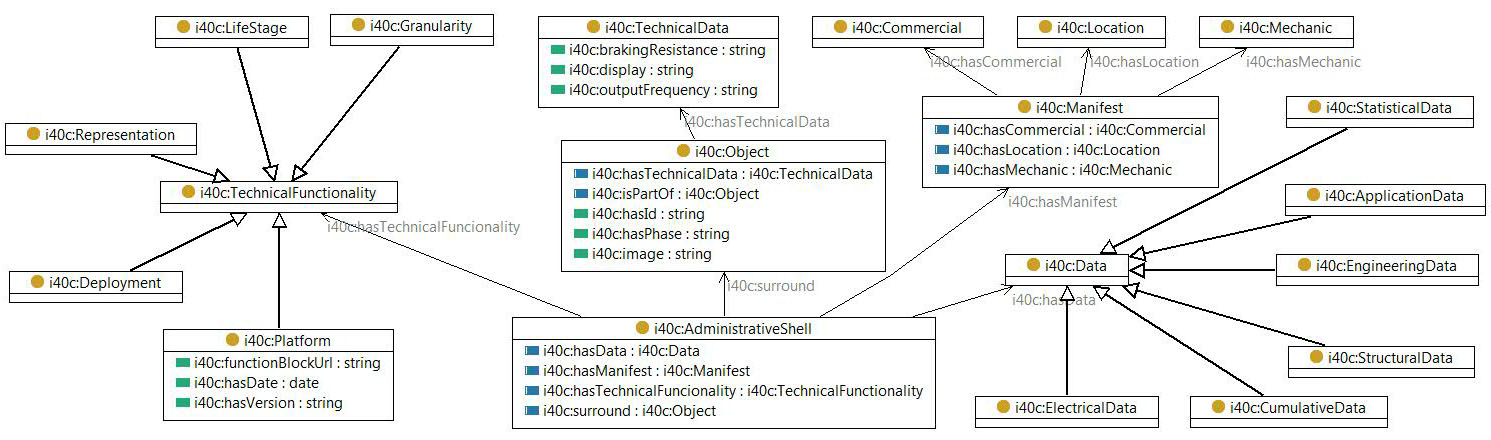}
  \caption{Overview of the core classes and relationships of the Administrative Shell Vocabulary.}
  \label{fig:AdminShellVoc}
\end{figure*}

In order to provide a semantic layer for the Administrative Shell we have developed an RDFS vocabulary.
This is depicted in \autoref{fig:AdminShellVoc} (for brevity reasons, we have omitted most of the properties and some of the classes).  
Our approach defines the Administrative Shell vocabulary by providing an ontological formalization of the terms and concepts presented in the I4.0 component. 
The central classes that cover the main concepts are \texttt{i40c:AdministrativeShell} and \texttt{i40c:Object}. 

The first class encapsulates the knowledge regarding the Administrative Shell.
This class wraps the object concept and connects it with most of the important features in the I4.0 domain.
The \texttt{i40c:TechnicalFunctionality} class allows to encode information regarding the version, url of the software that manage the I4.0 object.
Additionally, this class model information about the life stage, representation as well as the deployment of the mentioned object. 
Other important classes connected to the \texttt{i40c:AdministrativeShell} are the \texttt{i40c:Manifest} and \texttt{i40c:Data}.
Firstly, the information for commercial, mechanical  and location features  are modeled with the class \texttt{i40c:Manifest}.
Secondly, the class \texttt{Data} represents different types of data such as statistical, engineering, etc. 

The \texttt{i40c:Object} encapsulate the I4.0 object and important data related to it (e.g. identification, image, technical data).
In addition, information regarding the phases of the object and the entire life cycle are managed.
Given that the I4.0 object can be part of other components, we used the \texttt{isPartOf} Content Ontology Design pattern\footnote{\url{http://ontologydesignpatterns.org/wiki/Submissions:PartOf}} to capture this characteristic.

We argue that using a lightweight RDFS vocabulary, an important step towards realizing the I4.0 vision in real world applications is established. 

\section{Use case}
\label{usecase}

The vision of Industry 4.0 is centered around the concept of decentralized production and smart objects that participate in the production in terms of autonomy and decision-making.
To accomplish this goal, object metadata, data, and relations with other objects, need to be semantically described within the Administrative Shell. 
By doing so, the information provided by one object can be understood and exploited by other smart objects in the production chain.
To illustrate the applicability of our approach we detail, in this section, a use case where the semantic Administrative Shell is used to describe an I4.0 component and some of its basic relations.

\autoref{lst:UseCase} shows the semantic representation of the Administrative Shell for the \emph{Motor controller CMMP-AS-C2-3A-M3} object (a product of Festo AG\footnote{\url{https://www.festo.com/cat/en-gb_gb/products_CMMP_AS}}). 
For brevity, we describe here the most relevant data of the resources. 

\begin{figure}[tb]
  \centering    
  \includegraphics[width=1.0\linewidth]{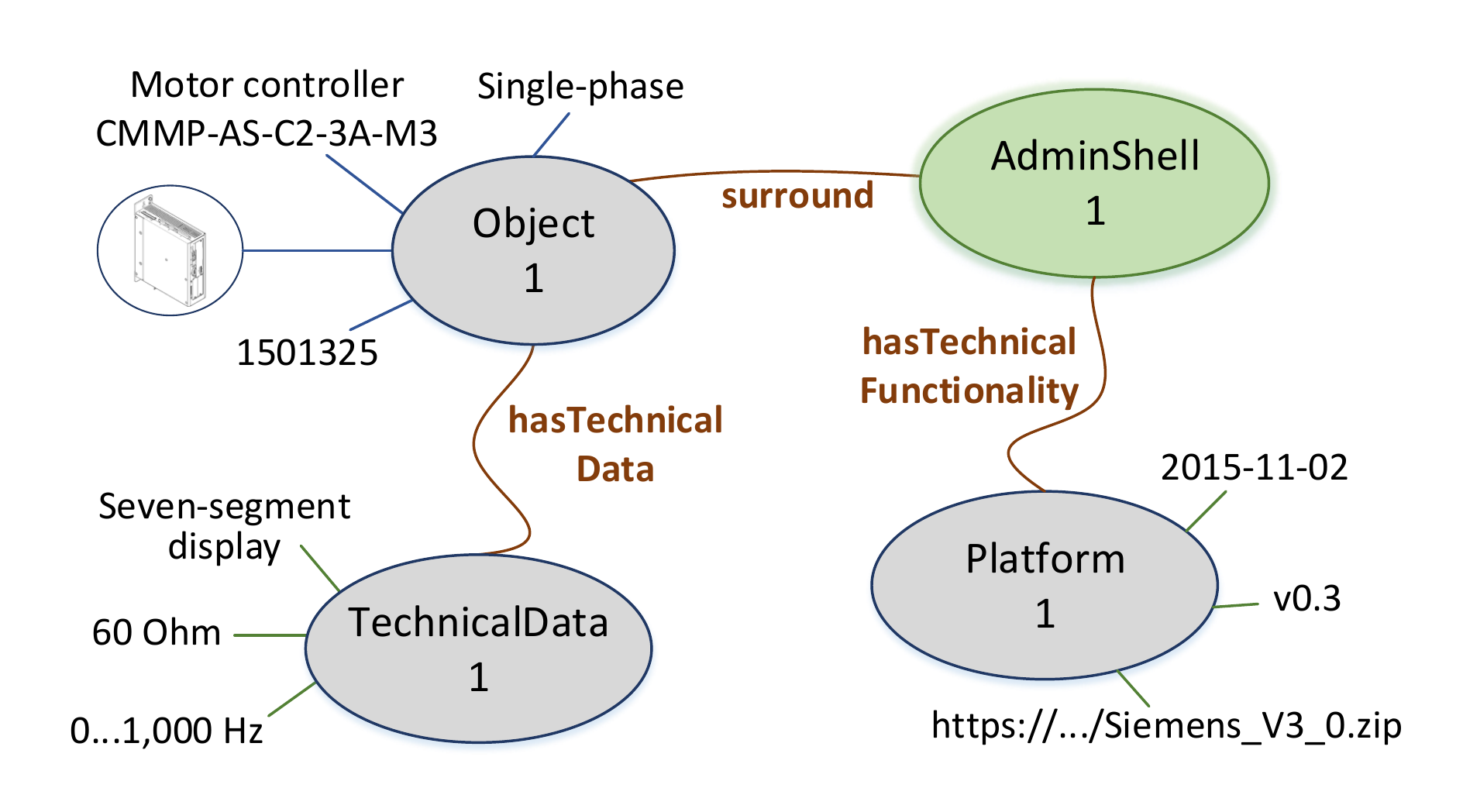}
  \caption{Example of Industry 4.0 objects graph}
  \label{fig:AdminShellUseCase}
\end{figure}

This example contains four instances of respective types.
An \texttt{AdminShell1} surround \texttt{Object1} and related it with the majority of the concepts in the domain, as the \texttt{Platform1} in this case. 
Also, \texttt{Object1} has its technical data defined on the resource \texttt{TechnicalData1} (cf. \autoref{fig:AdminShellUseCase}).

\begin{lstlisting}[style=turtle, caption={Semantic Administrative Shell for the servo motor controller CMMP-AS-C2-3A-M3}, label={lst:UseCase}]
@prefix i40c: <http://purl.org/eis/i40c/>.
@prefix rdf:  <http://www.w3.org/1999/02/22-rdf-syntax-ns#> .

i40c:AdminShell1      a     i40c:AdministrativeShell  ;
  rdfs:label                "AdminShell1"^^xsd:string ;
  i40c:surround             i40c:Object1 ;		
  i40c:hasTechFuncionality  i40c:Platform1 .

i40c:Object1          a     i40c:Object ;
  rdfs:label                "Motor control..."@en ;
  i40c:hasId                "1501325"^^xsd:string ;
  i40c:hasPhase             "Single-phase"@en ;
  i40c:hasTechnicalData     i40c:TechnicalData1 ;
  i40c:image                "<http://...b4dc.jpg>" .

i40c:TechnicalData1  a      i40c:TechnicalData ;
  rdfs:label                "TechnicalData1"@en ;
  i40c:BrakingResistance    "60 Ohm";
  i40c:Outputfrequency      "0...1,000 Hz" ;
  i40c:display              "Seven-segment display"@en . 
		
i40c:Platform1        a     i40c:Platform ;
  rdfs:label                "Function blocks ..."@en ;
  i40c:functionBlockUrl     "https://.../Siemens_V3_0.zip";
  i40c:hasDate              "2015-11-02"^^xsd:date ;
  i40c:hasVersion           "3.0"^^xsd:string .		

\end{lstlisting}

One of the main advantages of the Semantic Administrative Shell is the uniform data representation according to the RDFS model, which enables efficient integration and querying the data comprised in the shell.
In order to illustrate the data retrieval, we have designed simple SPARQL queries.
For example, it is relevant to know the technical characteristics of a component in its various phases (e.g. \emph{Single, Three}).
In the following query (cf. \autoref{lst:sparqlInitiated}), we construct an RDF graph that contains a description of the technical feature of an object with \emph{Single-phase}. 

\begin{lstlisting}[style=turtle, caption={SPARQL query to retrieve the electrical data with the state \emph{Initiated}}, label={lst:sparqlInitiated}]
@prefix i40c: <http://purl.org/eis/i40c/>.
@prefix rdfs: <http://www.w3.org/2000/01/rdf-schema#> .

CONSTRUCT {
  ?object         rdfs:label             ?name.
  ?technicalData  i40c:brakingResistance ?resistance .
  ?technicalData  i40c:outputFrequency   ?frequency .
} WHERE {
  ?object         i40c:hasPhase         "Single-phase"@en .
  ?adminShell     i40c:surround          ?object .
  ?object         rdfs:label             ?name.
  ?object         i40c:hasTechnicalData  ?technicalData .
  ?technicalData  i40c:brakingResistance ?resistance .
  ?technicalData  i40c:outputFrequency   ?frequency .
}
\end{lstlisting}

Another example is retrieving the information of the I4.0 component platform, during the maintenance cycle. 
The platform entities are referring to functional library elements, which are specific to a certain automation system.
The query modeled in \autoref{lst:platformData}, obtains the details of the platform like the name, version and the software URL that supports the I4.0 object.

\begin{lstlisting}[style=turtle, caption={SPARQL query to retrieve the version, date and URL of the platform software}, label={lst:platformData}]
@prefix i40c: <http://purl.org/eis/i40c/>.
@prefix rdfs: <http://www.w3.org/2000/01/rdf-schema#> .

CONSTRUCT {
  ?object       i40c:hasId                    ?objectId.
  ?platform     rdfs:label                    ?name .
  ?platform     i40c:hasVersion               ?version .
  ?platform     i40c:hasDate                  ?date .
  ?platform     i40c:functionBlockUrl         ?url .
} WHERE {
  ?adminShell   i40c:surround                 ?object .
  ?adminShell   i40c:hasTechnicalFuncionality ?platform .
  ?object       i40c:hasId                    ?objectId.
  ?platform     i40c:hasVersion               ?version .
  ?platform     i40c:hasDate                  ?date .
  ?platform     i40c:functionBlockUrl         ?url .
  ?platform     rdfs:label                    ?name .
}


\end{lstlisting}

The above use case shows how the Semantic Administrative Shell provides a more flexible data model.
This semantic representation helps to overcome the challenges that I4.0 is facing.

\section{Related Work}
\label{relatedWork}

\begin{table*}[tb]
	\centering
		\begin{tabular}{llllll}
		\hline
			\emph{Approach} & \emph{Basic Concept} & \emph{Identification} & \emph{Data model} & \emph{Organization Type} & \emph{Serialization} \\
		\hline	
			EDDL~\cite{eddl} & Device & Device and Unit ID & Object & Hierarchical & Text \\ \hline
			PML~\cite{brock2001physical} & Physical object & XML tag ID & Object & Hierarchical & XML \\ \hline
			OMM~\cite{haupert2011object} & Physical artifact & Primary ID and IDs Block & Element & Hierarchical & XML \\ \hline
			SPDO~\cite{janzen2008smart} & Product & URI/IRI & Resource & Hierarchical & OWL-DL\\ \hline
			Sem Admin Shell & Administrative Shell & URI/IRI & Resource & Hierarchical & RDF, RDF Schema, OWL\\ \hline
		\end{tabular}
	\caption{Comparison with related I4.0 component description approaches.}
	\label{tab:RelatedI40ComponentApproaches}
\end{table*}

Currently, there are some efforts discussing the need of bringing more semantics and data-driven approaches to I4.0.
~\cite{cheng2015semantic} presents guidelines, aiming to help on choosing the level of semantic formalization for the representation of the different types of I4.0 projects.
The crucial role of semantic technologies for mass customization is discussed in~\cite{wahlster2014semantic}.
This work recognizes semantic technologies as a glue to connect smart products, data and services. 
Obitko~\cite{obitko2015big} describes the application of semantics to I4.0 from the Big Data perspective. 
The features of Big Data, as well as, an ontology for sensor data are presented. 
\autoref{tab:RelatedI40ComponentApproaches} provides a comparison of our approach to the related I4.0 component description approaches.

\emph{Electronic Device Description Language} (EDDL) is a language to describe information related to digital components~\cite{eddl,conf/etfa/RundeWBS13}. 
Up to date, EDDL is available for a huge amount of devices that are currently utilized in the process industry.
EDDL is a text-based description file of the field device and its properties, which describes the data and how they should be displayed. 

The ~\emph{Object Memory Model} (OMM) is an XML-based format which allows for modeling of the information about individual physical elements~\cite{haupert2011object}.
In this work, the memory of the elements is partitioned to include different types of data regarding the identification, name, etc. 
The idea of this approach was to bring a semantic layer to the physical components but still suffers the intrinsic limitations of XML. 
However, it is envisioned, that elements in the OMM (so called blocks) contain RDF and OWL payload data.
Extending the concept of OMM \emph{Domeman}~\cite{haupert2013domeman} is a framework for representation, management, and utilization of digital object memories
The idea of bringing semantic descriptions of physical elements by combining OMM and a server realization has been conducted by~\cite{haupert2013object}.
Nevertheless, this work is focused on the identification of the elements and still rely on the above mentioned limitations of OMM format.

The \emph{Physical Markup Language }(PML)  is a common language for describing physical objects, processes and environments~\cite{brock2001physical}.
The goal of PML is to use this descriptions in remote monitoring and control of the physical environment.

Janzen~\cite{janzen2008smart} defines the Smart products as the connection of physical products and information goods which allow the embedding of digital product information into physical products.
In this approach the \emph{Smart Product Description Object} (SPDO) is presented. 
SPDO is a data model built on top of the \emph{DOLCE} ontology for describing smart products.
 
~\cite{bergweiler2015intelligent} defines an approach for distinguishing the local and global data structures stored in Active Digital Object Memories (ADOMe), smart labels with memory and processing capabilities. 
According to the author, this can be realized by storing the data in a unified structured format. 

The idea of bringing semantic descriptions of physical elements by combining OMM and a server realization has been conducted by~\cite{haupert2013object}.
Nevertheless, this work is focused on the identification of the elements and still rely on the above mentioned limitations of OMM format.

\section{Conclusion and Future Work}
\label{conclFutureWork}

In this paper, we have described an approach for semantically representing information about smart I4.0 devices with an Administrative Shell.
The approach is based on structuring the information using an extensible and light-weight vocabulary aiming to capture all relevant information.
Compared to prior approaches, the RDF-based Semantic Administrative Shell has a number of advantages.
The URI/IRI based identification scheme provides a unified way to identify all types of relevant entities, from physical objects, abstract concepts, properties, concrete raw and derived data, etc.
Existing standards (such as eClass, IEC device characteristics or AutomationML) can be more easily integrated and referenced.
Information about and from different objects can be easily integrated (since a basic integration can be achieved by merging sets of triples).
Accessing the information in a unified way is established by using SPARQL as query language.

We see this work as a first step in a larger research and development agenda aiming at equipping manufacturing equipment with semantics-based means for communication and data exchange.
In the medium to long term, we aim to bring more intelligence to the edge of production facilities thus promoting self-organization and resilience.

In future work, we envision to refine and expand the Administrative Shell vocabulary in order to provide support for a wide range of device types.
We intend to study the interaction between various devices equipped with Administrative Shells and research application scenarios, such as predictive maintenance.
Another interesting avenue of research is how Semantic Administrative Shells can be generated and populated from existing information systems and data sources available at the manufacturers.

\section*{Acknowledgments}
This work is supported by the German Ministry for Education and Research funded project LUCID and European Commission under H2020 for the project BigDataEurope (GA 644564).

\bibliographystyle{IEEEtran}
\bibliography{paper}

\end{document}